\documentclass[11pt,a4paper,longbibliography,aps,superscriptaddress]{revtex4-2}
\usepackage[top=2 cm, bottom=2 cm, left=2 cm, right=2 cm]{geometry}
\usepackage[utf8]{inputenc}
\usepackage{array}
\usepackage{amsmath}
\usepackage{amsfonts}
\usepackage{amssymb}
\usepackage{graphicx} 
\usepackage{romannum}
\usepackage{braket}
\usepackage{xcolor}

\begin{document}

\title{Tunable N-level EIT: Deterministic Generation of Optical States with Negative Wigner Function}

\author{Sutapa Ghosh}
\email{sutapa.g@campus.technion.ac.il, gad@ee.technion.ac.il, kaminer@technion.ac.il}
\affiliation{ECE department, Technion-Israel Institute of Technology, Haifa 32000, Israel}
\affiliation{University of Maryland, College Park, Maryland 20742, USA}
\author{Alexey Gorlach}
\affiliation{ECE department, Technion-Israel Institute of Technology, Haifa 32000, Israel}
\affiliation{Russel Berrie Nanotechnology Institute and Solid State Institute, Technion-Israel Institute of Technology, Haifa 32000, Israel}
\author{Chen Mechel}
\affiliation{ECE department, Technion-Israel Institute of Technology, Haifa 32000, Israel}
\affiliation{Russel Berrie Nanotechnology Institute and Solid State Institute, Technion-Israel Institute of Technology, Haifa 32000, Israel}
\author{Maria V. Chekhova}
\affiliation{ECE department, Technion-Israel Institute of Technology, Haifa 32000, Israel}
\affiliation{Max Planck Institute for the Science of Light, Erlangen, Germany}
\affiliation{Friedrich–Alexander Universität Erlangen–Nürnberg, Erlangen, Germany}
\author{Ido Kaminer}
\affiliation{ECE department, Technion-Israel Institute of Technology, Haifa 32000, Israel}
\affiliation{Russel Berrie Nanotechnology Institute and Solid State Institute, Technion-Israel Institute of Technology, Haifa 32000, Israel}
\affiliation{Diller Quantum Center, Technion-Israel Institute of Technology, Haifa 32000, Israel}
\author{Gadi Eisenstein}
\affiliation{ECE department, Technion-Israel Institute of Technology, Haifa 32000, Israel}
\affiliation{Russel Berrie Nanotechnology Institute and Solid State Institute, Technion-Israel Institute of Technology, Haifa 32000, Israel}
\affiliation{Diller Quantum Center, Technion-Israel Institute of Technology, Haifa 32000, Israel}

\begin{abstract}

 Strong optical nonlinearities are key to a range of technologies, particularly in the generation of photonic quantum states. The strongest nonlinearity in hot atomic vapors originates from electromagnetically induced transparency (EIT), which, while effective, often lacks tunability and suffers from significant losses due to atomic absorption.
 We propose and demonstrate an N-level EIT scheme, created by an optical frequency comb that excites a warm rubidium vapor. The massive number of comb lines simultaneously drive numerous transitions that interfere constructively to induce a giant and highly tunable cross-Kerr optical nonlinearity. 
 The obtained third-order nonlinearity values range from $1.2 \times 10^{-7}$ to $7.7 \times 10^{-7}$ $m^2 V^{-2}$. 
 Above and beyond that, the collective N-level interference can be optimized by phase shaping the comb lines using a spectral phase mask. Each nonlinearity value can then be tuned over a wide range, from 40\% to 250\% of the initial strength.
 We utilize the nonlinearity to demonstrate squeezing by self polarization rotation of CW signals that co-propagate with the pump and are tuned to one of the EIT transparent regions. Homodyne measurements reveal a quadrature squeezing level of 3.5 dB at a detuning of 640 MHz. When tuned closer to an atomic resonance, the nonlinearity is significantly enhanced while maintaining low losses, resulting in the generation of non-Gaussian cubic phase states. These states exhibit negative regions in their Wigner functions, a hallmark of quantum behavior. Consequently, N-level EIT enables the direct generation of photonic quantum states without requiring postselection.

\end{abstract}

\maketitle
\section{Introduction}
Optical nonlinearity is a cornerstone of numerous fundamental and applied fields of research~\cite{chang_14,seth_99,azuma_08,bartlett_02}. Decades of theoretical and experimental studies have provided a large variety of nonlinear effects such as second and third harmonic generation, spontaneous parametric down conversion (SPDC), four-wave mixing (FWM), generation of solitons, self focusing, and the Kerr effect~\cite{boyd_23}. In quantum technologies, nonlinear processes, particularly SPDC and FWM, are used to generate pairs of entangled photons,  which are essential for quantum communication and quantum key distribution~\cite{kwiat_95,bouwmeester_97}. 

Optical nonlinearities enable the generation of squeezed states of light~\cite{andersen_16}, which are imperative for precise measurements in quantum sensing and metrology, such as gravitational wave detection in interferometers like LIGO~\cite{aasi_13}. Squeezed light is so far the most efficient quantum state that has been created in the optical domain~\cite{andersen_16,glorieux_23}. The most popular processes to generate squeezed light are SPDC in $\chi^{(2)}$ materials~\cite{wu_86} as well as FWM and the Kerr effect in $\chi^{(3)}$ materials~\cite{mccormick_07,Heersink_05}. %A record squeezing level of 15 dB was demonstrated in an optical parametric oscillator with a periodically poled KTP crystal~\cite{henning_16}. The crystals used for this purpose have generally wide bandwidths but exhibit low nonlinear strengths, $\chi^{(2)}$ typically not exceeding 20 pm/V~\cite{pack_04,schiek_12}. These processes also require optimum phase matching, usually attained by temperature tuning or periodic polling. 

Compared to squeezed states, much more `exotic' are non-Gaussian quantum states, which are an extremely powerful resource for various quantum applications~\cite{brod_16}. In particular, they are a necessary ingredient for universal quantum gates~\cite{menicucci_06}. However, so far optical non-Gaussian states have only been produced using post-selection~\cite{kawasaki_24}. For example, an optical Schrödinger cat state was obtained by conditional homodyne detection of a Fock state~\cite{ourjoumtsev_07} or by subtracting a photon from a squeezed vacuum~\cite{ourjoumtsev_06}. Strong optical nonlinearities can potentially enable deterministic generation of non-Gaussian states. %as well as large squeezing levels, which has numerous applications in quantum optics, quantum computing, and communications. 
However, most materials have very weak intrinsic nonlinearities. For example,  typical values of third-order nonlinearity in glasses are on the order of $10^{-25}-10^{-17}$ m$^2$V$^{-2}$~\cite{Wang_14}. 

Strong nonlinearity can be generated in atomic systems. The electromagnetically induced transparency (EIT) process in atomic systems was used to generate a third order nonlinearity in cold atom systems placed in a magneto-optical trap~\cite{feizpour_15}, in Bose Einstein condensates with high atom densities~\cite{hau_99}, or in Rydberg systems with high dipole moments~\cite{sinclair_19}. All these systems show values of Re[$\chi^{(3)}$] ranging from $10^{-9}$ to $10^{-7}$ m$^2$V$^{-2}$, which is a few orders of magnitude higher than commonly found in the majority of materials. Furthermore, in atomic systems, FWM has been used extensively for deterministic squeezed light generation. Seeded FWM in a double lambda configuration was shown to produce 8 dB of twin beam squeezing between two output modes \cite{cormick_08}. Self polarization rotation (SPR) was also used often to generate squeezed light in atomic systems~\cite{ries_03,agha_10}. A squeezing level of 2.9 dB~\cite{barreiro_11} was measured in $^{87}$Rb vapor, which was enhanced through a magnetic field to 4.2 dB ~\cite{yu_22}. Despite all the advantages, atomic systems are strongly limited by atomic absorption, which induces significant losses and destroys the quantum states. This dictates to probe atoms far from their resonance, significantly decreasing the achievable nonlinear strength. Thus, atomic systems are promising but still lack the tunability and achievable nonlinear strength for integrating them into next-generation optical and quantum optical devices.

Here we report on achieving strong and tunable nonlinearities that generate deterministically not only squeezed states but also non-Gaussian states with negative Wigner functions. Specifically, we propose and demonstrate a nonlinear mechanism by which an optical frequency comb induces N-level EIT in warm rubidium vapor, generating a giant and highly tunable $\chi^{(3)}$ cross-Kerr nonlinearity. The N-level EIT process, induced by the comb, yields nonlinearities which are orders of magnitude stronger than those obtained in standard three-level EIT ~\cite{schmidt_96,bang_24,kang_03}. The strong nonlinearity is sensed by a CW probe that co-propagates with the pump and undergoes SPR. The
system we propose provides also an extremely wide tunability of the nonlinearity by controlling the comb properties through a phase mask or by exciting different N-level EIT resonances throughout the atomic spectrum. Our optimally achieved nonlinearity is $\mathrm{Re}~[\chi^{(3)}]=3.5 \cdot 10^{-7} ~\mathrm{m}^{2}/\mathrm{V}^{2}$ at a detuning of 640 MHz, which is increased to $\mathrm{Re}~[\chi^{(3)}]=7.7 \cdot 10^{-7} ~\mathrm{m}^{2}/\mathrm{V}^{-2}$ at a detuning of 70 MHz. We achieve this strong nonlinearity with low pump powers of less than $30~\mathrm{mW}$ by operating at small detunings (below $1~\mathrm{GHz}$), while utilizing the narrow transparency regions created by the N-level EIT. We quantify the nonlinearity strength through the quantum properties of the CW SPR signal, in particular through the degree of squeezing it experiences~\cite{agha_10,matsko_02}. A squeezed state with a squeezing level of $3.5~\mathrm{dB}$ having a $1~\mathrm{MHz}$ bandwidth is demonstrated. Finally, we generate a photonic non-Gaussian quantum state with negativity in its Wigner function without post-selection, as previously predicted theoretically in ~\cite{tyc_08}. This state resembles a cubic-phase state which is known to be a signature of a strong nonlinear interaction~\cite{kala_22,konno_21,llyod_99}, obtained here by the cross-Kerr interaction in $^{87}$Rb~\cite{tyc_08} at a detuning around 70 MHz. This state can be used as an ancillary state for implementing cubic phase gates, which play an essential role in quantum computing~\cite{takeda_19}. Cubic phase states have also been experimentally demonstrated in the microwave domain using superconducting qubits ~\cite{marti_23} and in the optical domain through measurement-based non-Gaussian operation on linear squeezed light~\cite{konno_21}. However, we, for the first time, {\it deterministically} generate cubic phase states in an atomic vapor system in the optical domain through the cross-Kerr interactions.
 
\begin{figure}[h]
\begin{center}
\includegraphics[width=\textwidth]{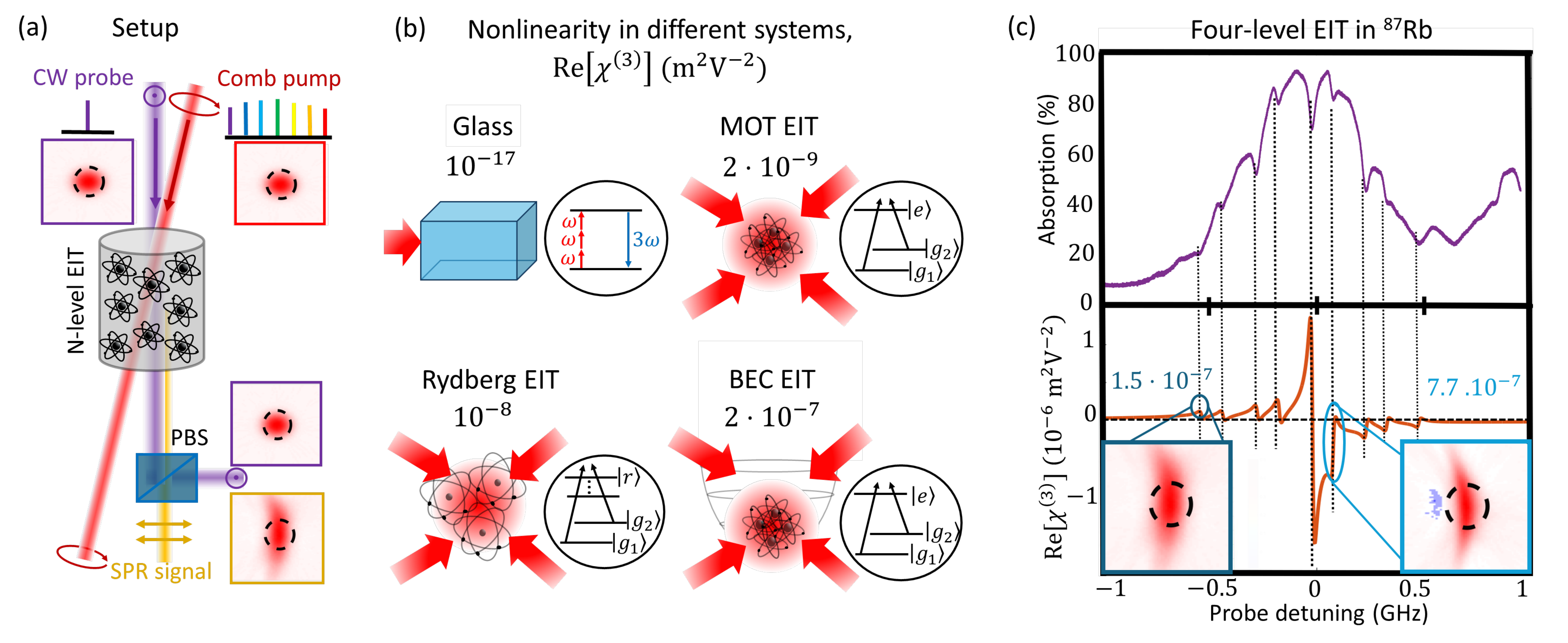}
\caption{{\bf Tunable strong third-order nonlinearity $\chi^{(3)}$ in warm rubidium vapor driven by a frequency comb.} \textbf{(a)} Schematics of the experimental set up. A frequency comb pump (red) and a CW laser probe (purple) interact with rubidium atoms in a vapor cell, causing N-level electromagnetically induced transparency (EIT) and the resultant nonlinearity. The probe light undergoes self-polarization rotation (yellow), which exhibits squeezing. We measure the squeezing level using a homodyne detection system and determine $\chi^{(3)}$ from the corresponding Wigner function. The Wigner functions of the frequency comb, the CW laser, and the signals after the polarizing beam splitter (PBS) are shown in small insets. \textbf{(b)} Comparison of the $\mathrm{Re}[\chi^{(3)}]$ generated by EIT in different systems~\cite{Wang_14,sinclair_19,hau_99,feizpour_15}. The thick red arrows represent laser beams used for optical cooling. \textbf{(c)} The absorption spectrum of the probe light passing through the rubidium cell. Measured first-order absorption is shown on top. The value of Re[$\chi^{(3)}$] is found from the measured squeezing level, and the corresponding Wigner functions, which are calculated based on the homodyne measurements of the SPR signal. For the non-Gaussian state, the nonlinearity is theoretically obtained for the particular used detuning. At resonance, the level of squeezing is limited by first-order absorption.}
\label{fig1}
\end{center}
\end{figure}

\begin{figure}[h]
\begin{center}
\includegraphics[width=\textwidth]{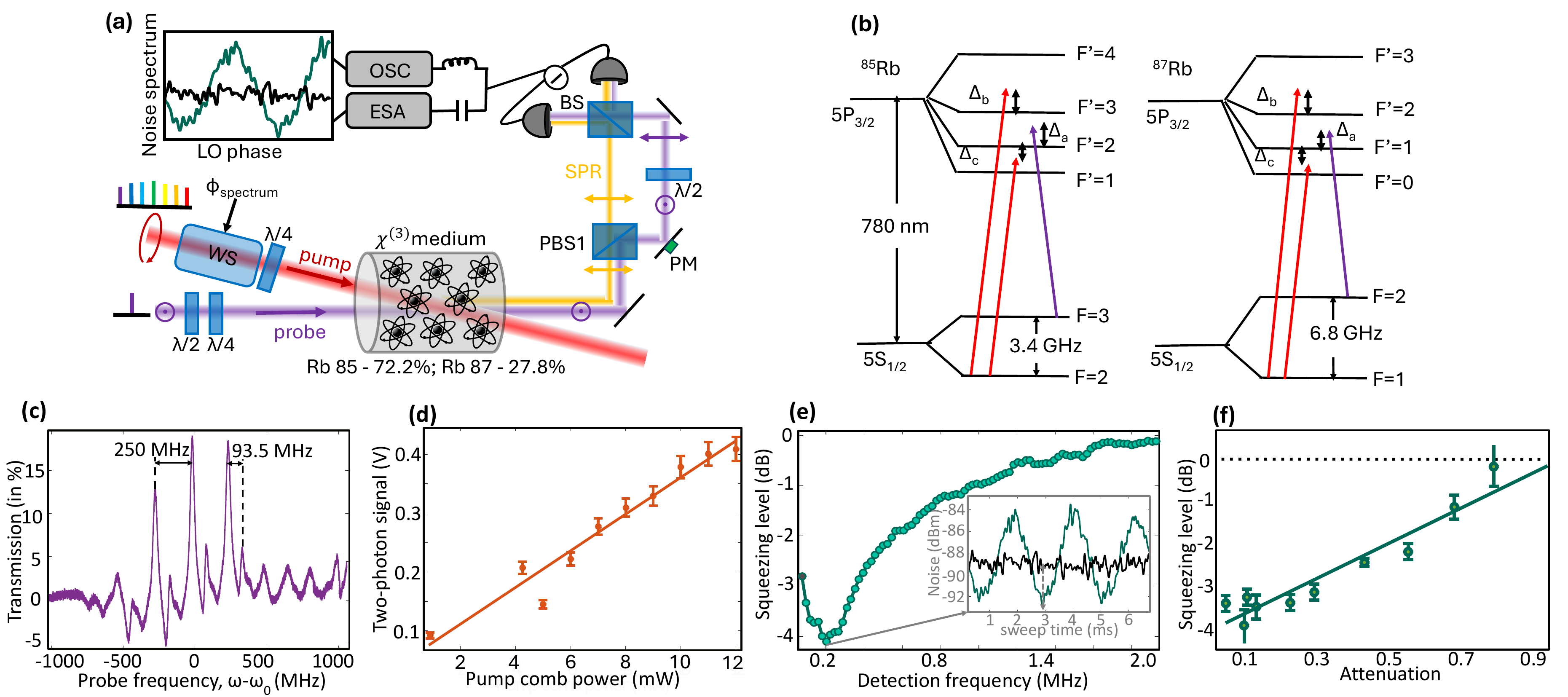}
\caption{\textbf{Giant third-order nonlinearity using N-level EIT.} \textbf{(a)} Experimental setup for N-level EIT. The pump frequency comb passes through a wave shaper (WS) that controls the phase spectrum of the comb and co-propagates with a CW signal in the rubidium vapor cell. A self polarization rotated (SPR) signal (yellow) is generated and is separated by a polarizing beam splitter from the strong classical component of the CW signal. The SPR signal is then mixed with the intense probe beam using a nonpolarizing beam splitter, and detected by balanced photodetection for homodyne measurement. \textbf{(b)} The energy level diagram of the N-level EIT excitation for $^{87}$Rb and $^{85}$Rb with $\Delta$ representing the detuning of the light fields from the corresponding transitions between atomic energy levels. \textbf{(c)} The transmission signal of the probe after the vapor cell shows two transmission peaks separated by $93.5~\mathrm{MHz}$ which repeat every $250~\mathrm{MHz}$ (the comb repetition rate). The Doppler broadened absorption spectrum is subtracted from the transmission data. The profile of the transmission spectrum becomes asymmetric as a function of detuning. \textbf{(d)} The EIT resonance strength increases linearly with the comb laser power, verifying that the system is far from saturation. \textbf{(e)} The frequency of the probe laser is tuned to one of the EIT resonance peaks and the noise spectrum at the output of the balanced detection system is measured as a function of the local oscillator (LO) phase, as shown in the inset. The black curve shows the shot noise level, measured when the SPR signal is blocked, while the green line shows the squeezed signal. The squeezing level as a function of detection frequency is plotted indicating that the bandwidth of the squeezed light is 1 MHz, determined by the narrow EIT resonance spectrum. \textbf{(f)}  The squeezing level is measured as a function of linear attenuation in the squeezed signal path. As the attenuation increases, the squeezing decreases linearly and eventually converges to the shot-noise limit (dashed line).}
\label{fig2}
\end{center}
\end{figure}

\section{Results}
\subsection{Giant third-order susceptibility using N-level EIT}
Figure~\ref{fig1}(a) depicts the scheme we propose. We illuminate a cell of warm rubidium vapor at $60 ^\circ$C with a circulary polarized optical frequency comb serving as a pump and a co propagating, linearly polarized CW probe. The comb is $1~\mathrm{THz}$ wide with a repetition rate of $250~\mathrm{MHz}$, containing some $4000$ lines. Our scheme uses electromagnetically induced transparency (EIT) to make the medium transparent within several narrowband windows. Each narrow EIT region modifies the optical response of the medium, enabling effects such as slow light propagation and giant cross-Kerr nonlinearity with few photons~\cite{fleishauer_05}. The multiple comb lines induce interference between many transition pathways, creating N-level EIT, leading to transmission peaks with enhanced $\chi^{(3)}$ nonlinearity~\cite{schmidt_96,bang_24}. This nonlinearity induces SPR on part of the CW probe field, which is then isolated from the original probe field through a polarizing beam splitter. SPR is known to exhibit non-classical properties, such as squeezing~\cite{agha_10,matsko_02} and sub-Poissonian photon statistics~\cite{tanas_79}. The nonlinearity is determined from the degree of squeezing of the SPR signal which has a simple relationship to $\mathrm{Re}[\chi^{(3)}]$ ~\cite{agha_10,matsko_02}. We obtain the quantum properties of the SPR signal using a homodyne measurement of the SPR signal where the non polarization rotated part of the probe serves as a local oscillator (LO). The phase of the LO is scanned through a piezo-mounted mirror placed in its path.

The vapor cell contains a mixture of rubidium isotopes $^{85}$Rb ($72\%$) and $^{87}$Rb ($28\%$). Figure~\ref{fig2}(b) shows the energy level diagram of both isotopes, from which we extract the number of EIT levels and the expected nonlinearity. For $^{87}$Rb (right), two consecutive comb lines couple the ground state $F=1$ to two excited states $F'=1,2$ while the probe couples the second ground state $F=2$ to the two excited states $F'=1,2$. This creates interference between the transitions from two consecutive comb lines and the probe, leading to four level EIT with two transparency peaks that are separated by $93.5$ MHz, as shown in the transmission spectrum, Fig.~\ref{fig2} (c). The $250$ MHz repetition rate of the comb dictates the periodicity of the transparency patterns and can be tuned over the atomic spectrum in the range of $\pm 1~ \mathrm{GHz}$. For $^{85}$Rb (Fig.~\ref{fig2}(b) left), the energy levels are very close to each other, which enables coupling of multiple rubidium energy levels to two consecutive comb lines, leading to EIT with more than four levels~\cite{paspalakis_02}. In particular, we observe EIT with $N=7$ levels if the comb induces transitions $F=2 \rightarrow F'=1,2,3$ and EIT with $N=6$ levels if it induces transitions $F=3 \rightarrow F'=2,3,4$.

The lineshapes of the transparency peaks in Fig.~\ref{fig2}(c) are formulated using a generalized Lorentzian function~\cite{finkelstein_23}
\begin{equation} \label{eqn_2}
a(\delta) = \gamma \frac{A\gamma + B(\delta-\delta_0)}{\gamma^2+(\delta_0-\delta)^2 +C},
\end{equation}
where $\delta=\Delta_c-\Delta_a$ is the two-photon detuning ($\delta=0$ at resonance) and $\gamma$ is the EIT resonance width. The constants $A, B 
\text{ and }C$ depend on the one-photon detuning, $\Delta$. $\delta_0$ is a fitting parameter introduced to account for the shift in the resonance spectral position from the exact two-photon resonance condition~\cite{mikhailov_04}. $\delta_0$ is a function of the one-photon detuning $\Delta$. As $\Delta$ increases, the shape of the transmission resonance changes while maintaining the two-photon resonance condition ($\delta=0$). 

Figure~\ref{fig2}(d) shows the strength of the EIT transparency peaks as a function of pump power, demonstrating a linear dependence and indicating that the system is far from saturation under the power range used in the present experimemnt.

To estimate the $\mathrm{Re[\chi^{(3)}]}$, we calculate the atomic population transfer under N-level EIT using the Bloch equations~\cite{paspalakis_02}. For example, under four-level EIT, the density matrix element corresponding to the probe transition is 
\begin{equation}\label{eqn_1}
(\rho_{13})_{4\mathrm{L}} = \frac{\Omega_a/2}{(\Delta_a -i\gamma_{13}) + \frac{\Omega_c^2/4}{(\Delta_c-\Delta_a+i\gamma_{12}) - \frac{\Omega_b^2/4}{(\Delta_c-\Delta_a-\Delta_b+i\gamma_{14})}}},
\end{equation}
where $\Omega_a$, $\Omega_b$, and $\Omega_c$ are the Rabi frequencies of the probe and the two consecutive comb lines, which are detuned from the atomic transitions by $\Delta_a$, $\Delta_b$ and $\Delta_c$ respectively; $\gamma_{13}$ and $\gamma_{14}$ are the relaxation rates of the excited levels while $4\mathrm{L}$ denotes four-level EIT. We calculate the theoretical $\text{Re}[\chi^{(3)}]$ value by Taylor expansion of Eq.~\ref{eqn_1}\begin{equation}\label{eqn_3}
\text{Re}[\chi^{(3)}]_{4\mathrm{L}} = \frac{N|\mu_{13}|^2|\mu_{24}|^2}{2\varepsilon_0\hbar^3} \frac{1}{\Omega_c^2\Delta_b},
\end{equation}
where $N$ is the number density of the atoms in the vapor cell, $\mu_{13}$ and $\mu_{24}$ are the dipole matrix elements for the corresponding transitions, $\hbar$ is the reduced Plank's constant, and $\varepsilon_0$ is the vacuum permittivity. The four level EIT configuration we obtain is predicted to enhance $\text{Re}[\chi^{(3)}]$ relative to a three-level EIT by six orders of magnitude within the experimental parameters used here. The enhancement increases as the detuning widens. A similar enhancement, by nine orders of magnitude, was predicted in ~\cite{schmidt_96}. We similarly calculate $\text{Re}[\chi^{(3)}]_{\mathrm{NL}}$ for $N=6$ and $7$ and find respectively four and five transparency peaks, which repeat at the comb repetition rate. N-level EIT configurations with $N$ larger than $4$ show a more moderate enhancement of the nonlinearity compared to $N=4$. The frequency of each EIT peak can be calculated from Eq.~\ref{eqn_1} as shown in Figure~\ref{fig2}(c).

Figure~\ref{fig2}(e) shows the squeezing level as a function of detection frequency in $^{85}$Rb under the $F=2 \rightarrow F'= 1,2,3$ transitions at a detuning of $640\pm28~\mathrm{MHz}$. The bandwidth of the squeezed light is 1 MHz, dictated by the narrow EIT resonance spectrum width $\gamma=30 \pm 5 ~\mathrm{MHz}$. The inset shows typical noise levels as a function of the LO phase (here shown for $200~\mathrm{kHz}$). When the quantum path is blocked, there is only the noise of the balanced photodetector output (black curve), which determines the shot-noise level. The quantum SPR light exhibits noise reduction below the shot-noise level at particular LO phases, as shown by the green curve. Figure~\ref{fig2}(f) shows the squeezing level as a function of linear attenuation, which was placed manually into the path of squeezed light verifying the convergence of the squeezed state to the shot-noise level at maximum attenuation. 

For $^{87}$Rb, we measured a squeezing level of $1.8 \pm 0.3~\mathrm{dB}$ at a one-photon detuning of $470 \pm 20$ MHz with $20 \%$ absorption, corresponding to $\text{Re}[\chi^{(3)}] = (1.5 \pm 0.4) \cdot10^{-7}$ $\mathrm{m}^2\mathrm{V}^{-2}$. Theoretically, for such detuning, we get $\text{Re}[\chi^{(3)}]$ to be $(1.5 \pm 0.1) \cdot 10^{-7}$ $\mathrm{m}^2\mathrm{V}^{-2}$. For $^{85}$Rb under the transitions $F=3 \rightarrow F'= 2,3,4$, we obtained a squeezing level of $2.2 \pm 0.5$ dB and a $\text{Re}[\chi^{(3)}]=(3.5 \pm 1.5) \cdot10^{-7}$~$\mathrm{m}^2\mathrm{V}^{-2}$ at a detuning of $370 \pm 25~\mathrm{MHz}$ with $30\%$ absorption, compared to a theoretical value of $\text{Re}[\chi^{(3)}]=(2.4\pm0.1) \cdot10^{-7}\mathrm{m}^2\mathrm{V}^{-2}$. 
The maximum squeezing level we obtained for $^{85}$Rb was $3.5 \pm 0.2 $ dB under the transitions $F=2 \rightarrow F'= 1,2,3$ at a detuning of $640 \pm 30~\mathrm{MHz}$ and with $10\%$ absorption. This corresponds to the nonlinearity of $\text{Re}[\chi^{(3)}]=(1.5 \pm 0.2) \cdot10^{-7}$~$\mathrm{m}^2\mathrm{V}^{-2}$ compared to a theoretical value of $\text{Re}[\chi^{(3)}]=(2.4 \pm 0.1) \cdot10^{-7}\mathrm{m}^2\mathrm{V}^{-2}$. The difference between the two isotopes is mainly due to the different vapor pressures of the atoms, which determines the number density $N$. The measurements were performed at EIT resonances that are spectrally placed far from the atomic resonances in order to reduce the absorption.

\subsection{Effect of the frequency comb phase on the nonlinearity}
\begin{figure}[h]
\begin{center}
\includegraphics[width=\textwidth]{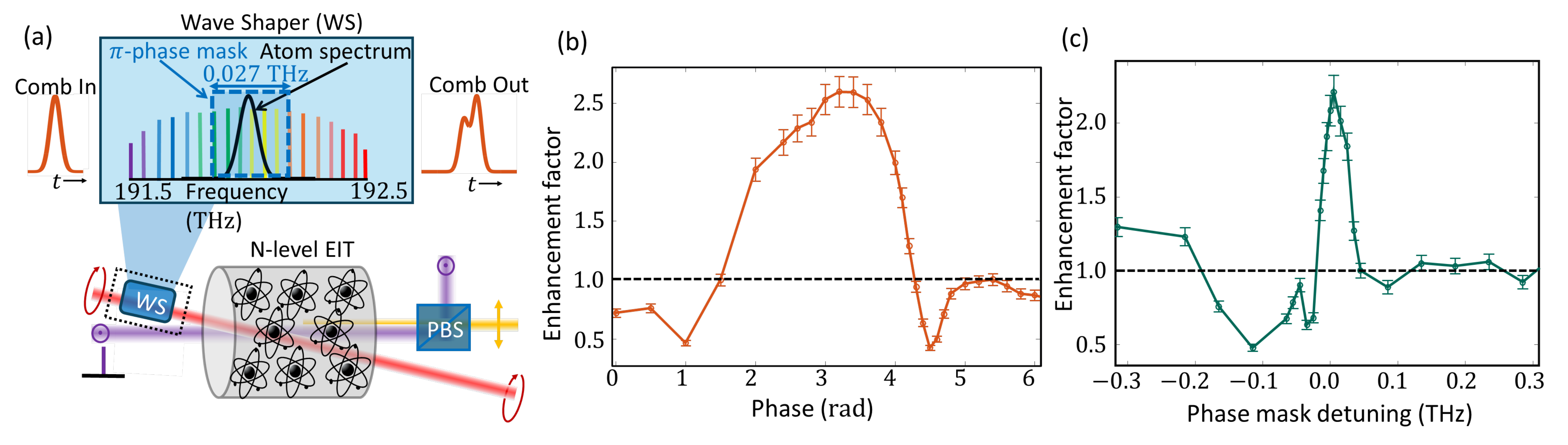}
\caption{{\bf Tunability of the nonlinearity of the comb-driven-vapor.} The multi-frequency comb signal, which induces multiple N-level EIT processes,  is phase shaped, providing tunability of the nonlinearity by adding up constructively all the N-level EIT processes. \textbf{(a)} A constant phase is added at the atom resonance frequency. \textbf{(b)} The nonlinearity enhancement factor as a function of the phase added to 110 comb lines by the $0.027$ THz wide phase mask. \textbf{(c)} Enhancement factor as a function of the center position of the optimized phase mask relative to the resonance frequency.}
\label{fig3}
\end{center}
\end{figure} 

The scheme we describe implements N-level EIT by coupling multiple atomic energy levels to consecutive comb lines.  Furthermore, a single N-level EIT process has numerous contributions from the entire comb spectrum. Although the contributions of the off-resonant comb lines are weaker than those of the resonant lines, their cumulative effect adds significantly to the nonlinearity. Moreover, controlling the relative phase between these contributions allows tuning the EIT process and the resultant nonlinearity. To this end, we employ a phase mask on the input comb spectrum. Figure~\ref{fig3}(a) shows the modified experimental setup. We use a waveshaper (WS) to imprint a uniform phase mask on the comb signal within a spectral range of $0.027~\mathrm{THz}$ (blue dashed rectangle) with a tunable central frequency. The WS bandwidth covers roughly 110 lines with a resolution of 1 GHz, namely four lines.

Figure~\ref{fig3}(b) shows the nonlinearity enhancement factor as a function of the phase added by the WS relative to the unmasked comb (black line) where the center of the phase mask is tuned to the atomic resonance. With this particular phase pattern, we demonstrate nonlinearity enhancement of $2.5$ for a phase of $(2n+1)\pi$ and a reduction by a factor of $2.5$ when the added phase is $(2n+1) \pi/2$, with $n$ being an integer. The experimental data exhibit a maximum at a phase of $\pi$ and a minimum at $0.64 \pi/2$ instead of $\pi/2$. However, the minimum is at the expected phase value of $3\pi/2$. The phase mask enables therefore a $6$-fold increase of the dynamic range of the nonlinearity strength. Theoretically, the transition amplitude corresponding to the off-resonant optical signal adds a $\pm \pi/2$ phase, symmetrically detuned around the resonance~\cite{stowe_08,yoon_00}. Therefore, a phase addition of $\pi$ leads to a constructive contribution of the comb lines, as observed. 

Figure~\ref{fig3}(c) shows the nonlinearity enhancement as a function of the central frequency of the phase mask spectrum relative to the atomic resonance frequency. The value of the phase mask is kept at $\pi$ with the width of $0.027$ THz. The black line shows the enhancement value without the phase mask. The effect of the phase mask is seen within the range of approximately $\pm 0.3$ THz. The maximum enhancement has been naturally observed with the center of the phase mask at atomic resonance.  

\begin{figure}[t]
\begin{center}
\includegraphics[width=\textwidth]{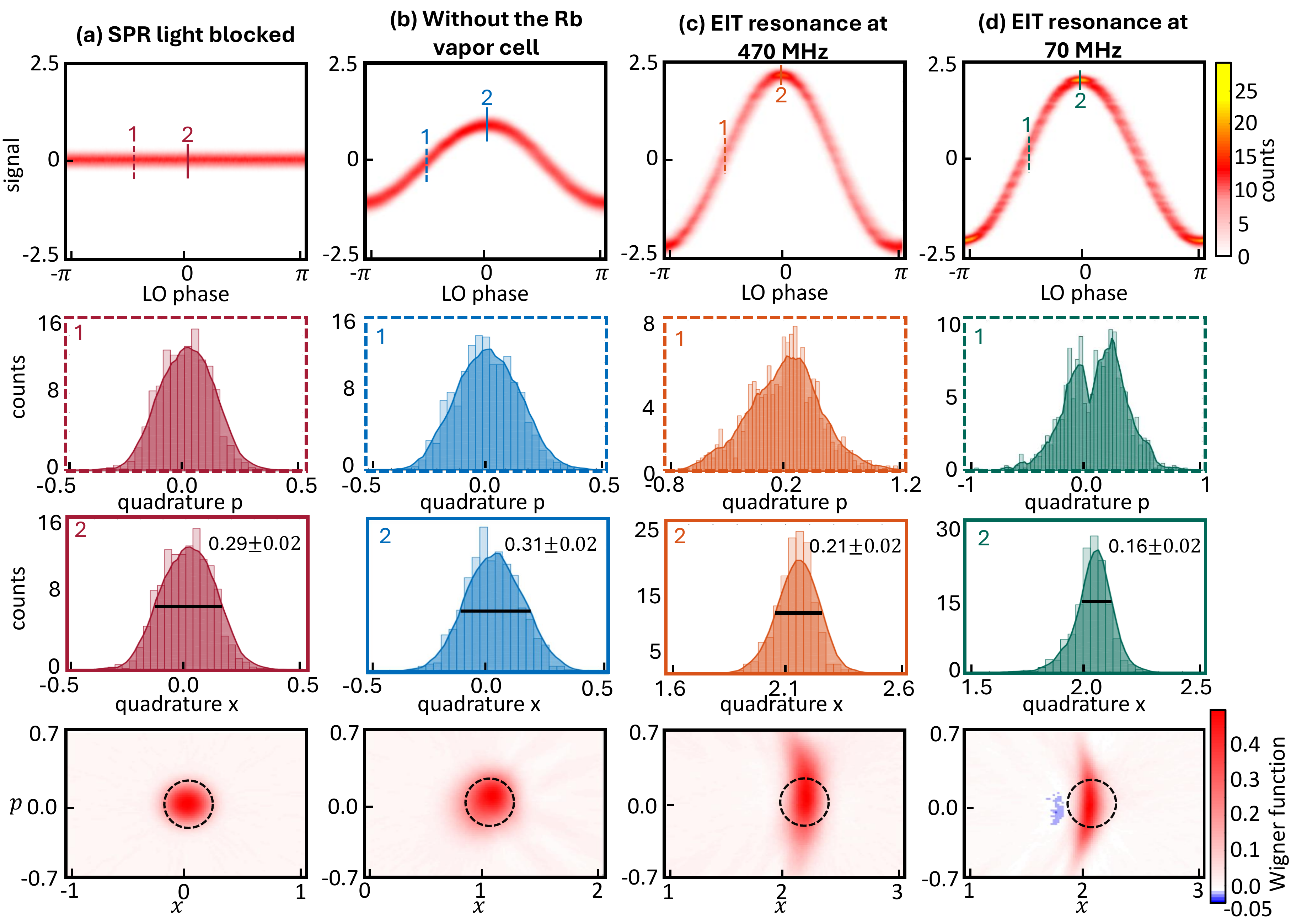}
\caption{{\bf Characterization of the measured quantum light.} States of the self-polarization rotated (SPR) signal for various probe field detunings. \textbf{(a)} Vacuum state, measured when the SPR signal is blocked. \textbf{(b)} A coherent state measured for a weak probe that does not pass through the rubidium cell. \textbf{(c)} A quadrature squeezed state measured by tuning the probe to one of the EIT peaks with a moderate nonlinearity, $\text{Re}[\chi^{(3)}]= 1.2 \cdot 10^{-7}\text{m}^2\text{V}^{-2}$. \textbf{(d)} A non-Gaussian state obtained by tuning the probe to an EIT line that is close to the atomic resonance, thereby exhibiting a much stronger nonlinearity, $\text{Re}[\chi^{(3)}]= 7.7 \cdot 10^{-7}\text{m}^2\text{V}^{-2}$. The top panel represents the signal of the balanced homodyne detector as a function of the LO phase. The second and third rows show the signal histograms obtained at two different phases, corresponding to the probability distributions of two different quadratures. The bottom panel contains the corresponding Wigner functions. For both the squeezed state and the non-Gaussian state, the histogram width is reduced, which reveals amplitude squeezing, as clearly seen in the third row. The Wigner function of the non-Gaussian state (d) has negative parts, which is responsible for the double peaked histogram in the second row. In (c), the Wigner function and, correspondingly, the left histogram are slightly distorted relative to a Gaussian profile. In all Wigner functions, the dashed circle shows the vacuum uncertainty, for reference.}
\label{fig4}
\end{center}
\end{figure} 

\subsection{Deterministic generation of states with Wigner-function negativities}
Our scheme utilizes cross-Kerr nonlinearity to achieve squeezing of the probe signal, from which we deduce the strength of the nonlinearity. We furthermore used the same effect and setup to demonstrate deterministic generation of a non-Gaussian state at optical frequencies. We used  homodyne detection to characterize the SPR signal and reconstructed the correspoinding Wigner functions~\cite{breitenbach_97}. We demonstrated that for a sufficiently strong nonlinearity, which is achieved through probing the EIT resonance close to the atomic resonance, the states exhibit negativity in some parts of the Wigner function ~\cite{tanas_79}. 

We reconstruct Wigner functions of different light states from different EIT resonances within the spectrum of $^{87}$Rb atoms. Figure~\ref{fig4} describes four distinct cases. The top panels show the raw data, namely the measured current distributions as  functions of the local oscillator phase. The second and third rows show the histograms of the signal at two selected phases, corresponding to the probability distributions of the $p$ and $x$ quadratures. The bottom panels show the Wigner functions, calculated using the inverse Radon transform~\cite{smithey_93} from the top panels, with a dashed black circle indicating the vacuum state as a reference. Figure~\ref{fig4}(a) displays the vacuum state obtained by blocking the SPR signal. Figure~\ref{fig4}(b) shows a coherent state obtained when the rubidium cell is bypassed while using using a weak probe. Figure~\ref{fig4}(c) describes a $1.8~\mathrm{dB}$ squeezed state, obtained at a detuning of  $470~\mathrm{MHz}$, corresponding to $\text{Re}[\chi^{(3)}]=1.2 \cdot10^{-7}~\mathrm{m}^2\mathrm{V}^{-2}$. Compared to the vacuum state, the state is narrower in the $x$ quadrature, and the $x$-quadrature histogram is asymmetric as shown in the third row. Figure~\ref{fig4}(d) represents a non-Gaussian state whose Wigner function exhibits negtivity, obtained at a detuning of $70~\mathrm{MHz}$ where the nonlinearity is $\text{Re}[\chi^{(3)}]=7.71 \cdot10^{-7}$~$\mathrm{m}^2\mathrm{V}^{-2}$. The Wigner function has a crescent (i.e., banana-like) shape that closely resembles a cubic phase state, as expected due to the strong $\chi^{(3)}$ nonlinearity~\cite{tyc_08}. The double-peaked $p$ quadrature histogram of Fig.~\ref{fig4}(d) (second row) is evidence for the Wigner function's negativity~\cite{MacRae_12}. The overall detection efficiency of the homodyne system is $72\%$, which is higher than the $50 \%$, threshold to detect negativity in a Wigner function ~\cite{nugmanov_22}. These cubic phase states are attractive candidates for nonlinear phase gates in continuous-variable quantum computing schemes~\cite{andersen_15,takeda_19}.

\section{Discussion and outlook}
The system we introduce here combines several factors to achieve a giant cross-Kerr nonlinearity. We utilize N-level EIT using the narrow spaced energy levels of rubidium atoms and a frequency comb. Additionally, we optimize the comb phase pattern for constructive interference between many transition pathways. This combination enables to achieve a strong nonlinearity together with low losses and a very wide range of tunability.

A key advantage of our scheme is the high ratio of nonlinearity to absorption. We utilize large detuning from atomic resonances to lower absorption yet the system yields strong nonlinearities due to the numerous number of contributing comb lines. The main factor that limits further reduction of absorption is the moderate EIT transparency level, which is around 25$\%$ in warm atoms. It has been shown however to improve by at least 80$\%$ for cold atoms~\cite{kang_03}. By optically cooling the atoms, the process of generation and detection of quantum state can become much more robust. The nonlinearity can be further enhanced by exciting Rydberg states, whose dipole moments are larger by approximately three orders of magnitude~\cite{sinclair_19}. In this case, the nonlinearity can be increased by six orders of magnitude.

We extract the Re[$\chi^{(3)}$] values from the squeezing level of the SPR signal. The main source of error in this method is the instability of the free running probe laser. The linewidth of the probe laser is around 100 kHz but it drifts randomly around the resonance by 20 MHz during the total measurement time.  This drift is comparable to EIT resonance width, which is 30 MHz. These errors can be overcome by phase-locking the probe laser and the comb. 

We have demonstrated the generation of a non-Gaussian state of light with a negativity in the Wigner function. To the best of our knowledge, this is the first deterministic generation of quantum light states with a negative Wigner function in the optical range. Other demonstrations are in the microwave range ~\cite{marti_23} or using post selection ~\cite{konno_21}. The non-Gaussian state we measured resembles the cubic phase state, which holds potential for universal quantum computation~\cite{takeda_19} and for generating other complex quantum states~\cite{he_23}. 

In the next step, this scheme must overcome several challenges. The effect of absorption loss is still non-negligible, causing decaying or dephasing of the quantum state~\cite{kala_22}. The losses inhibit probing the resonance region in $^{85}$Rb, where the nonlinearity is stronger than in $^{87}$Rb. Near atomic resonances, the enhanced nonlinearity can be probed at significantly lower optical powers, potentially extending to the single-photon regime. Thus, if the losses are overcome, N-level EIT will enable nonlinear interactions on a single-photon level which is a longstanding challenge in quantum optics. Based on our experimental parameters, the phase shift imprinted on the probe field at the single-photon level is $400~\mathrm{\mu rad}$ at $370 ~\mathrm{MHz}$ detuning, which can be larger for lower detunings, compared to previously reported values of $13-300~\mathrm{\mu rad}$~\cite{vivek_13}.

Looking forward, we envision our scheme to advance research in several directions. N-level EIT can be used as a resource for single-photon nonlinearity. It provides new tuning capabilities to generate photonic quantum states as well as to store photon states for quantum repeater application~\cite{sangouard_11}. The squeezing level obtained here could be used as a resource for quantum metrology and quantum memories by utilizing acousto-optic modulators to scan them across the resonance. Finally, our method is a good candidate to generate non-Gaussian states for universal quantum gates in continuous-variable quantum computing~\cite{andersen_15}.

\section{Methods}
\subsection{Experimental Setup}
The experimental system comprises two lasers, shown in Fig. 2(a). One is a commercial fiber-based frequency comb laser from Menlo systems, with a repetition rate of $250 ~\mathrm{MHz}$ centered at $1560 \mathrm{nm}$, serving as a pump. The second is a free-running tunable external-cavity CW diode laser acting as the probe. The repetition rate of the comb is RF locked to GPS with a stability of $2 \cdot 10^{-13}$ $s^{-1}$. A detailed overview of the comb laser system and its stabilization units are given in~\cite{DCS_23}.  Following amplification by EDFAs, the pump and probe signals are frequency doubled to $780\, \mathrm{nm}$ using periodically poled lithium niobate crystals. The pump laser passes through a wave shaper (Finisar 1000S) that controls the phase spectrum of the comb. The wave shaper has a bandwidth of 27 GHz and its resolution is 1 GHz. The pump beam then passes through the rubidium cell, which is heated to $60^{\circ}\mathrm{C}$, at a small angle with respect to the probe beam in a co-propagating configuration. The rubidium cell contains two isotopes $^{85}$Rb and $^{87}$Rb with concentrations of $72.2\%$ and $27.8\%$ respectively. The vapor cell is surrounded by three layers of $\mu-$metal sheets to shield it from any external residual magnetic field. The probe signal passes through two polarizing beam splitters, which separate the SPR signal from the intense probe. The strong probe passes through a half-wave plate and is used as a local oscillator (LO) in a homodyne measurement. The phase of the LO is tuned by applying a voltage to a piezo-mounted mirror. It is combined with the polarization-rotated signal on a non-polarizing beam splitter and detected by two photodetectors. The two photocurrents are subtracted from each other with the resulting signal passing through a bias tee that feeds the DC signal to an oscilloscope and the AC part to an electrical spectrum analyzer where the noise spectrum is characterized.

\section{Acknowledgements}
This work was partially funded by the Qbit program within the Hellen Diller Quantum Center at Technion. SG acknowledges the financial support of the Diller Quantum Center. AG  acknowledges financial support from the Russel Berrie Nanotechnology Institute, the Diller Quantum Center and the Azrieli Fellowship. CM acknowledges the financial support of the Diller Quantum Center and the Azrieli Fellowship.

\def\bibsection{\subsection*{\refname}} 
\bibliography{Arxiv_Manuscript}

\end{document}